\documentclass[sigconf,nonacm]{acmart}

\AtBeginDocument{%
  }

\setcopyright{acmlicensed}
\copyrightyear{2025}
\acmYear{2025}
\acmDOI{XXXXXXX.XXXXXXX}

\acmConference[Conference acronym 'XX]{Make sure to enter the correct
  conference title from your rights confirmation emai}{June 03--05,
  2018}{Woodstock, NY}
\acmISBN{978-1-4503-XXXX-X/18/06}

\begin{document}

\title{MIRACL-VISION: A Large, multilingual, visual document retrieval benchmark}


\author{Radek Osmulski*}
\affiliation{%
  \institution{NVIDIA}
  \city{Brisbane}
  \country{Australia}
}
\email{rosmulski@nvidia.com} 

\author{Gabriel de Souza P. Moreira*}
\affiliation{%
  \institution{NVIDIA}
  \city{S\~ao Paulo}
  \country{Brazil}}
\email{gmoreira@nvidia.com}

\author{Ronay Ak*}
\affiliation{%
  \institution{NVIDIA}
  \city{Sarasota}
  \country{USA}}
\email{ronaya@nvidia.com}  

\author{Mengyao Xu*}
\affiliation{%
 \institution{NVIDIA}
 \city{Santa Clara}
 \country{USA}}
\email{mengyaox@nvidia.com}  

\author{Benedikt Schifferer}
\authornote{All authors contributed equally to this research.}
\affiliation{%
  \institution{NVIDIA}
  \city{Berlin}
  \country{Germany}
}
\email{bschifferer@nvidia.com}

\author{Even Oldridge}
\affiliation{%
  \institution{NVIDIA}
  \city{Vancouver}
  \country{Canada}}
\email{eoldridge@nvidia.com}

\renewcommand{\shortauthors}{Osmulski et al.}

\begin{abstract}
Document retrieval is an important task for search and Retrieval-Augmented Generation (RAG) applications. Large Language Models (LLMs) have contributed to improving the accuracy of text-based document retrieval. However, documents with complex layout and visual elements like tables, charts and infographics are not perfectly represented in textual format.
Recently, image-based document retrieval pipelines have become popular, which use visual large language models (VLMs) to retrieve relevant page images given a query. Current evaluation benchmarks on visual document retrieval are limited, as they primarily focus only English language, rely on synthetically generated questions and offer a small corpus size. Therefore, we introduce MIRACL-VISION\footnote{The dataset is available at https://huggingface.co/datasets/nvidia/miracl-vision}, a multilingual visual document retrieval evaluation benchmark. MIRACL-VISION covers 18 languages, and is an extension of the MIRACL dataset, a popular benchmark to evaluate text-based multilingual retrieval pipelines. MIRACL was built using a human-intensive annotation process to generate high-quality questions. In order to reduce MIRACL-VISION corpus size to make evaluation more compute friendly while keeping the datasets challenging, we have designed a method for eliminating the "easy" negatives from the corpus. 
We conducted extensive experiments comparing MIRACL-VISION with other benchmarks, using popular public text and image models. 
We observe a gap in state-of-the-art VLM-based embedding models on multilingual capabilities, with up to 59.7\% lower retrieval accuracy than a text-based retrieval models. Even for the English language, the visual models retrieval accuracy is 12.1\% lower compared to text-based models. MIRACL-VISION is a challenging, representative, multilingual evaluation benchmark for visual retrieval pipelines and will help the community build robust models for document retrieval. 
\end{abstract}



\maketitle

\section{Introduction}

\begin{figure}[ht]
    \centering
    \includegraphics[width=1.0\linewidth]{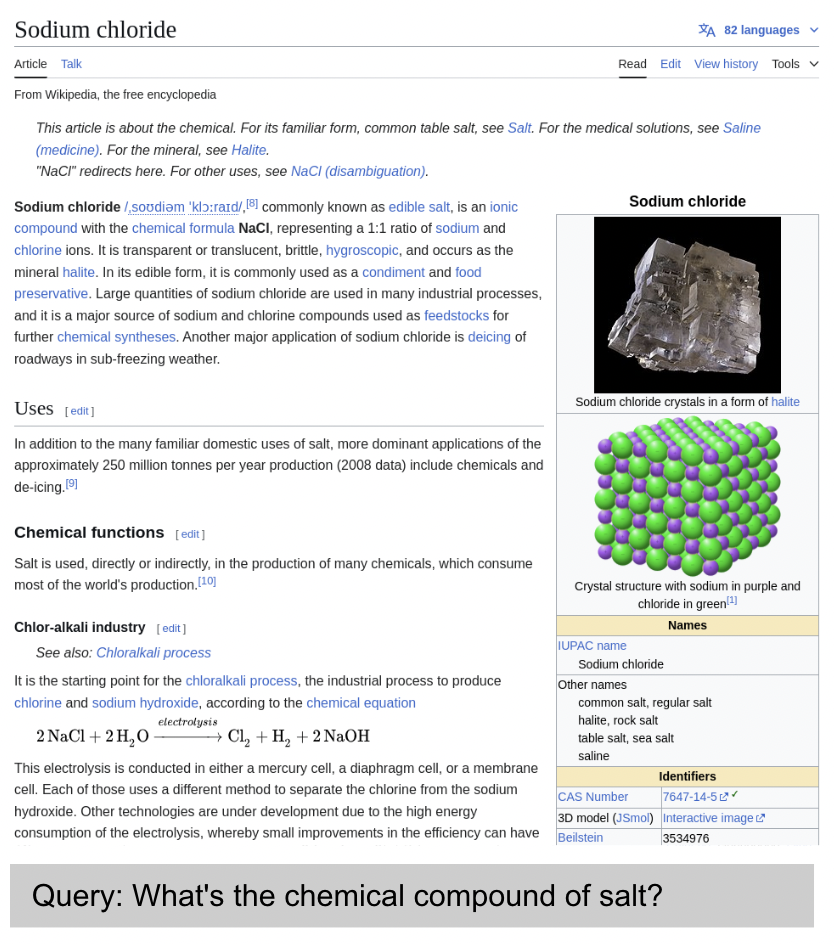}
    \caption{Example of a User Query and Document Image of MIRACL Vision}
    \label{fig:example_1}
\end{figure}

Retrieval-Augmented Generation (RAG) has become a popular approach to provide context for Large Language Models, enabling LLMs to answer zero-shot questions, e.g., about content that was not seen during training.

Many companies have been adopting RAG to create assistants that leverage their internal documents - like reports, contracts, presentations - to improve their customer service or increase productivity and quality of their internal processes.

\begin{figure}[ht]
    \centering
    \includegraphics[width=1.0\linewidth]{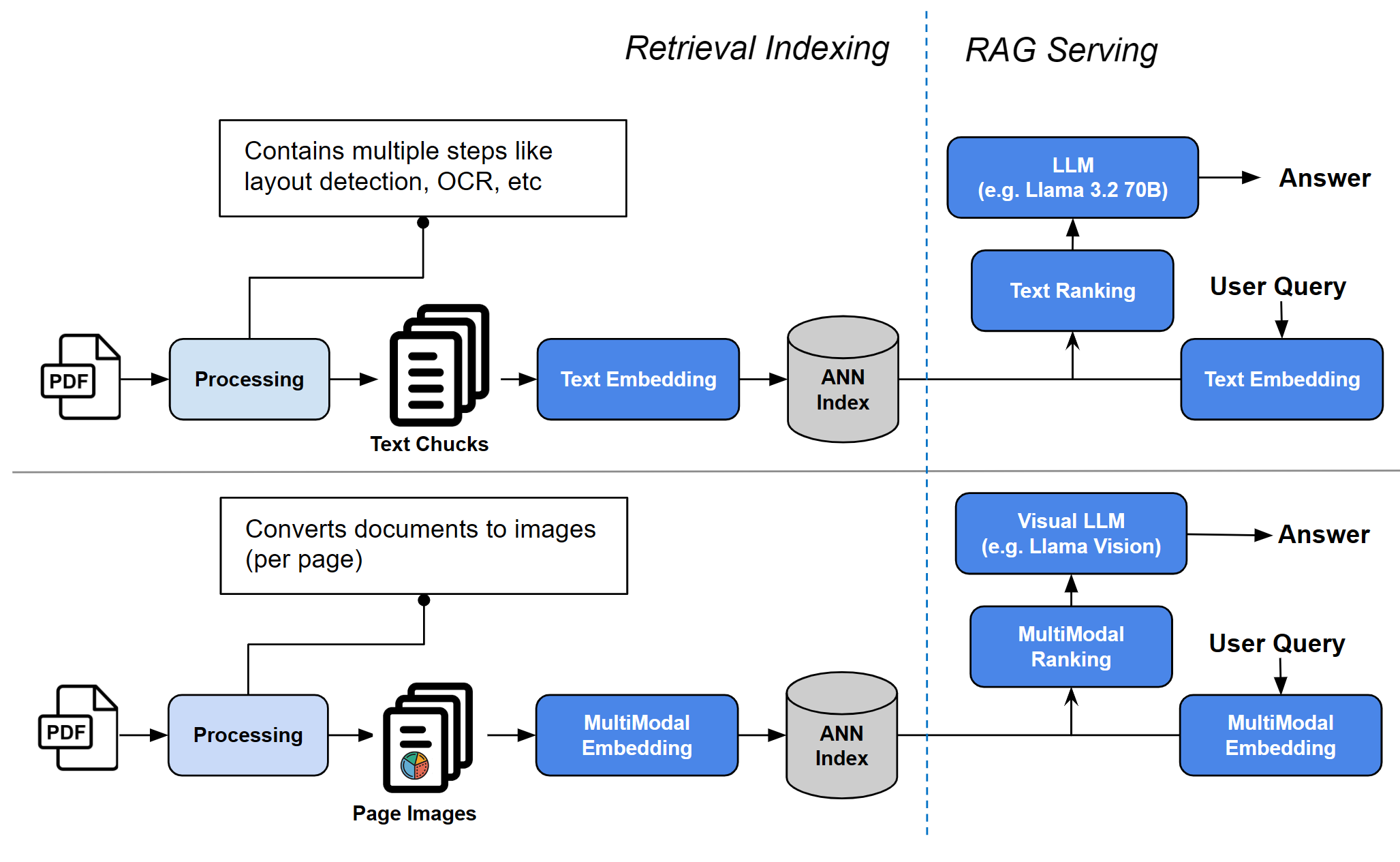}
    \caption{Visualization of a text-based and image-based RAG pipeline}
    \label{fig:pipeline}
\end{figure}

A key component of RAG applications is retrieval. In a typical text-based retrieval pipeline, documents need to be first parsed for text extraction, which is split into chunks that are embedded for dense retrieval. 
Older scanned documents are represented as images and require Optical Character Recognition (OCR) to extract text. Most modern document formats store the actual text and avoid the need of OCR, but more complex document layouts (e.g. two-column documents, text interleaved with images and tables) make text extraction more challenging. This text-based retrieval scenario demands non-trivial ingestion and indexing pipelines for documents, which might involve specialized models. For example, document layout detection models to segment the page elements, usage of LLMs to caption figures and tables in natural language, a chunking strategy that is aware of the structure of the document.

A recent approach has been to represent pages as images\cite{ma2024unifying} and retrieve them using Visual LLMs (VLMs), which have built-in OCR capabilities. VLMs are generation models capable of taking both text and images as input. 

VLMs have been adapted as multimodal constrastive embedding models, that can align images and text representation in a shared embedding space. Recent VLM-based document retrieval models have been released, such as DSE-Qwen2\cite{ma2024unifying}, GME-Qwen2\cite{zhang2024gme}, ColPali and ColQwen\cite{faysse2024colpaliefficientdocumentretrieval}.

Some benchmarks have been introduced to evaluate VLM-based embedding models capability to retrieve document pages: ViDoRe\cite{faysse2024colpali}  and VDR\cite{vdrmultilingual}. However, they have some limitations, like questions generated synthetically, small and not challenging document corpus, and lack of multi-language coverage.

In this paper, we introduce the MIRACL-VISION benchmark, which is focused on evaluating the multilingual support of visual document retrieval. 

MIRACL-VISION is based on MIRACL, a popular benchmark for multilingual text-based retrieval, including high-resource (e.g. English) and low-resource languages (e.g. Swahili), and languages with non-Latin alphabets (e.g., Arabic, Japanese, Korean, Russian). MIRACL authors invested a significant effort to collect representative user questions from Wikipedia articles by native speakers. 

In MIRACL-VISION data collection process, we leverage the high-quality MIRACL questions about Wikipedia articles in multiple languages, generate the corresponding images of the first page of those articles, and filter the dataset to reduce its size while keeping hard-negatives / distractors for retrieval evaluation.

The major contributions of this paper are summarized as follows:

\begin{itemize}
    \item The release of MIRACL-VISION, a comprehensive benchmark for evaluation of multilingual visual document retrieval and its comparison with existing benchmarks;
    \item We describe the data collection process for MIRACL-VISION, which can be adapted to create visual retrieval versions of other text retrieval datasets based on documents or webpages;
    \item We provide a benchmark of state-of-the-art visual document embedding models on multilingual retrieval task with MIRACL-VISION and compare them with text-based embedding models on an equivalent text-based dataset.
\end{itemize}

We believe MIRACL-VISION will be helpful for the community to evaluate the multilingual capabilities of vision-based retriever pipelines.

\section{Background}

In this section, we discuss related work on retrieval benchmarks and VLM-based constrastive embedding models.

\subsection{Text retrieval benchmarks}

Machine Learning benchmarks are important to help the community to set targets for real-world problems and tasks, gauge research progress towards those goals over time, and provide a common ground to compare different methods and models.

One of the main benchmarks for text information retrieval is BEIR\cite{thakur2021beir}. It is a selection of 18 English retrieval datasets from 9 heterogeneous retrieval tasks, including Question Answering retrieval datasets - NQ, HotpotQA and FiQA-2018 - that are relevant for RAG applications.

MIRACL \cite{zhang2022making} is a multilingual benchmark for text retrieval, comprising 18 different languages that cover over three billion native speakers around the world. Their authors leveraged native speakers to generate around 77k queries and evaluate top-k query-passage pairs produced by a retrieval system. This careful human-annotation has been very valuable for multilingual text retrieval evaluation. Since there is no comprehensive multilingual benchmark for visual document retrieval, as we discuss in the next section, we introduce MIRACL-VISION in this work.

\subsection{Visual document retrieval benchmarks}

The Visual Document Retrieval Benchmark (ViDoRe)\cite{faysse2024colpali} is popular for evaluating page-level document retrieval. It covers many document types (e.g., financial reports, administrative and medical documents, academic papers, among others) and features questions about different visual elements (text, tables, charts, infographics). ViDoRe adapts 6 existing Visual Question Answering (VQA) datasets for retrieval, and create other 5 datasets by generating questions using a proprietary VLM from corpuses of documents. The majority of its datasets are in English, only two of them cover French language.

VDR-Multilingual\footnote{https://huggingface.co/datasets/llamaindex/vdr-multilingual-test}\cite{vdrmultilingual} is another benchmark for visual document retrieval. It covers English, French, German, Italian, and Spanish languages. For each language and category of questions (text, visual, mix) there are 100 questions and a corpus 1000 page images. A VLM was used to generate synthetic questions that were human-reviewed.

We compare our MIRACL-VISION with ViDoRe and VDR-Multilingual benchmarks
in Section~\ref{sec:comparison_visual_benchmarks}

\subsection{VLM-based embedding models}

Contrastive dense embedding models represent variable-length information as a fixed dimension vector that can be used for downstream tasks, like retrieval. 
Transformer models have been fine-tuned to serve as text embedding models using encoder-based architectures (DPR\cite{karpukhin2020dense}, E5 \cite{wang2022text}), and decoder models (E5-Mistral \cite{wang2023improving}). Multilingual text embedding models have been released, like 
\textit{multilingual-e5-large} \cite{wang2024multilingual}, \textit{snowflake-arctic-embed-l} \cite{yu2024arctic}, \textit{bge-m3} \cite{bge-m3}, and \textit{gte-multilingual-base} \cite{zhang2024mgte}.

Visual LLM models  (e.g. PaliGemma\cite{steiner2024paligemma}, SmolVL\cite{smolvlm}, QwenVL\cite{QwenVL}, and Eagle2\cite{li2025eagle}) combine vision and language capabilities, enabling tasks like image captioning, question answering, and multimodal retrieval.

VLMs typically integrate a vision encoder model (e.g. SigLIP) with a Language model (e.g. Llama) by using a connector (e.g. MLP) that projects and aligns the text and image embedding spaces.

VLMs can be adapted from a generation model into and multimodal embedding model by pooling the Transformer embedding outputs and training it with constrastive learning, bringing together in the embedding space the positive text-image pairs, e.g., matching a question with the corresponding document page image that contains the answer.

Some recent representative VLM-based models for visual document retrieval are \textit{dse-qwen2-2b-mrl-v1}\cite{ma2024unifying}, \textit{gme-Qwen2-VL-2B-Instruct}\cite{zhang2024gme},  \textit{vdr-2b-multi-v1}, and \textit{colqwen2-v1.0}\cite{faysse2024colpaliefficientdocumentretrieval}. In Section~\ref{sec:vlms_benchmark}, we evaluate those models on visual document retrieval benchmarks, including our MIRACL-VISION.

\section{Methodology}

In this section, we describe our design and process for extending MIRACL and generating the MIRACL-VISION data set to benchmark multilingual visual document retrieval (Section~\ref{sec:extension}). Afterwards, we compare the characteristics of MIRACL-VISION with other datasets in Section \ref{sec:stats}.


\subsection{MIRACL-VISION generation process}
\label{sec:extension}

\begin{figure*}[ht]
    \centering
    \includegraphics[width=0.9\linewidth]{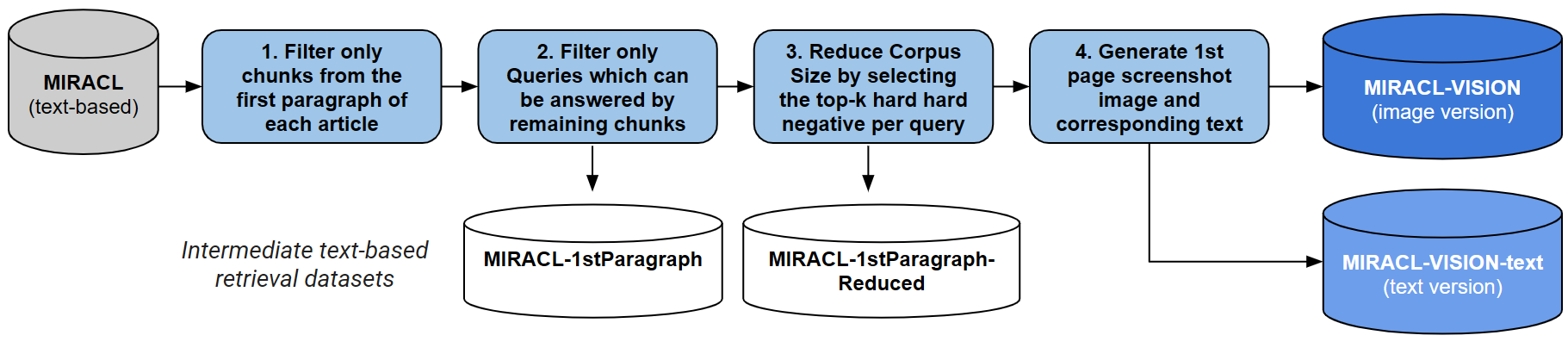}
    \caption{Visualization of the process to create MIRACL-VISION and the intermediate datasets}
    \label{fig:collection}
\end{figure*}

We illustrate our general process to extend MIRACL to MIRACL-VISION in Figure~\ref{fig:collection}, which is inspired by the construction of the \textit{Wiki-SS} dataset \cite{ma2024unifying}

To generate MIRACL-VISION, we have designed a process that reuses MIRACL human-generated questions and replaces the Ground Truth (GT) annotated text passages, i.e., that contain the answer, by an image of the document where the GT passage is contained. That process also involves steps to reduce dataset size while keeping hard-negatives for all questions. We detail the steps in this section.

\textbf{Step 1. Filtering 1st Paragraph per Article} \newline
MIRACL corpus was extracted from Wikipedia articles, that can be long. For that reason, they are split into multiple chunks to keep their length manageable to be embedded. For example, English Wikipedia has ~5.7M articles and the corpus of English MIRACL has ~32M chunks.

To be able to reuse MIRACL questions for visual retrieval, we had to design a process to locate the chunk containing the answer within an article and extract the corresponding document page image containing that chunk. We did not find a reliable solution to extract images from chunks in any part of the document. We simplified the process by keeping only the first chunk. That way, we can always take the first page of a Wikipedia article, as it is ensured to contain the first paragraph.


\textbf{Step 2. Selecting Answerable Questions} \newline
After we removed all chunks which are not the first paragraph, some questions did not have the corresponding GT anymore in the corpus. In this step, we removed all the questions that do not have any positive document in the corpus. We name this intermediate dataset as \textbf{MIRACL-1stParagraph}.

\textbf{Step 3. Reducing Corpus Size while keeping it challenging for evaluation} \newline
As described in Table~\ref{tab:miracl_stats}, some languages still have a large corpus after keeping only the first paragraph chunks, such as English with 5.7M documents / chunks. 
For a large number of documents, extracting document images and running the evaluation pipeline would be costly and require significant computational resources, making it impractical as an evaluation dataset for the research community. Besides that, most documents of a large corpus are 
irrelevant to annotated queries, as it is easy to distinguish them from the correct documents.

We have designed a method to reduce the dataset size, while keeping its hardness for retrieval evaluation. It only keeps documents in the corpus that are either positives or hard negatives for at least one question.
To perform that filtering, we use the \textit{multilingual-e5-large} text embedding model to embed all questions and documents, compute the cosine similarity among those embeddings to get the top-k (top-100 for English and top-50 for other languages) most similar documents to the question and only keep them in the corpus. The resulting corpus is much smaller, but still challenging for retrieval, as it keeps the main distractor documents for each query. We name this intermediate dataset as \textbf{MIRACL-1stParagraph-Reduced}.

\textbf{Step 4. Generating Image and Text} \newline
MIRACL is based on Wikipedia, a publicly available website with user-friendly terms and conditions. We follow a similar process as described in \cite{ma2024unifying}. For each document in MIRACL, we download the corresponding Wikipedia article. We modify the HTML code to render only relevant content,  removing some elements like sidebar, and header. Then we extract an image of the first vertical 2048 pixels of the article with Playwright\footnote{https://playwright.dev/}, crop it to 980px x 980px pixels and save it to disk. You can see examples in  Figure~\ref{fig:example_1} and in Appendix~\ref{sec:more_examples} of user queries with corresponding Wikipedia article containing the answer in the first paragraph. 

In addition, we extract the text from the HTML body, keeping the first 12 sentences\footnote{Based on manual inspections, the extracted Wikipedia article images have approximately 12 sentences.} as an approximate text representation of the extracted image. We name this dataset as \textbf{MIRACL-VISION-text}.

Appendix~\ref{sec:more_examples} provides an example of extracted image and corresponding text from a Wikipedia article. 


\subsection{Comparison of retrieval benchmark datasets}
\label{sec:stats}

In this section, we can compare statistics of MIRACL-VISION with other evaluation datasets.

\subsubsection{Statistics of MIRACL and MIRACL-VISION}

Table \ref{tab:miracl_stats} provides the statistics of the MIRACL, MIRACL-1stParagraph and MIRACL-VISION. The original MIRACL has an average ~5.9M text chunks per language and filtering on 1st paragraph reduces the corpus size to an average of ~1M chunks. As some queries are not answerable with the filtered dataset, the average number of queries per language is reduced from 750 to 439. 

A corpus of 1M images per language would require significant computation for evaluating models. Therefore, in Step 3 in Section \ref{sec:extension} we describe how we reduce the corpus size to an average of 18,819 documents by removing chunks that are not relevant to any question, i.e. we only keep hard-negatives / distractors. This approach ensures correlated retrieval results while reducing corpus size and speeding up evaluation.

\subsubsection{Comparison of visual document retrieval benchmarks}
\label{sec:comparison_visual_benchmarks}

We compare the characteristics of MIRACL-VISION with other popular vision benchmarks in Table \ref{tab:comparision_benchmarks}. ViDoRe provides 8 English and 2 French datasets; vdr-multilingual contains English, French, German, Italian, and Spanish datasets and MIRACL-VISION has a total of 18 different languages, including low-resource languages or languages with non-Latin alphabets. The number of queries per language dataset are for those datasets (300-483).
The average corpus size per language of MIRACL-VISION is  6x larger than the other datasets. 

One limitation of MIRACL-VISION is that the queries are mainly based on information from the text, whereas ViDoRe and vdr-multilingual datasets contain queries for tables, charts and infographics. However, we believe that text-based queries are highly relevant to evaluate the multilingual capabilities for vision-based models. 

The researchers that prepared MIRACL dataset have trained native speakers to ask relevant questions given a "prompt" passage, where the prompt passage cannot answer the question but the question will likely be answered by the remaining text and verified it later. 

ViDoRe and Vdr-multilingual generated synthetic queries for some datasets with LLMs and manually reviewed them. In our experience, prompting an LLM to formulate questions given a document has the tendency that specific keywords will be repeated in the questions. Therefore, the questions might not be representative of open queries from a user who is not biased by a specific document.

ViDoRe repurposed existing Visual Question-Answering (VQA) datasets as retrieval datasets. One particular issue of that approach is that they may contain queries that are specific to a sentence, table or image, and might not make sense for retrieval. For example, ViDORe's docvqa test set contains question like \textit{"What is the table number?"} or \textit{What is the email address provided?}, which makes sense to ask a VLM when the image is provided, but doesn't make sense for document retrieval.

\begin{table*}[htb]
\caption{Comparison of number of queries and number of documents between original MIRACL, MIRACL filtered on 1st paragraph and MIRACL-VISION.}
\begin{tabular}{l|ll|ll|ll}
                  & \multicolumn{2}{c}{MIRACL (original)}             & \multicolumn{2}{c}{MIRACL-1stParagraph}        & \multicolumn{2}{c}{MIRACL-VISION}                 \\
\textbf{Language} & \textbf{\# of queries} & \textbf{\# of document chunks} & \textbf{\# of queries} & \textbf{\# of documents} & \textbf{\# of queries} & \textbf{\# of documents} \\ \hline
Arabic            & 2896                   & 2061414                  & 2127                   & 656982                   & 2127                   & 75444                    \\
Bengali           & 411                    & 297265                   & 229                    & 63762                    & 229                    & 8495                     \\
Chinese           & 393                    & 4934368                  & 189                    & 1246389                  & 189                    & 8672                     \\
English           & 799                    & 32893221                 & 447                    & 5758285                  & 447                    & 42971                    \\
Farsi             & 632                    & 2207172                  & 342                    & 857827                   & 342                    & 15846                    \\
Finnish           & 1271                   & 1883509                  & 791                    & 447815                   & 791                    & 33679                    \\
French            & 343                    & 14636953                 & 142                    & 2325608                  & 142                    & 6990                     \\
German            & 305                    & 15866222                 & 129                    & 2651352                  & 129                    & 6302                     \\
Hindi             & 350                    & 506264                   & 184                    & 148107                   & 184                    & 8004                     \\
Indonesian        & 960                    & 1446315                  & 603                    & 446330                   & 603                    & 23842                    \\
Japanese          & 860                    & 6953614                  & 387                    & 1133444                  & 387                    & 17909                    \\
Korean            & 213                    & 1486752                  & 130                    & 437373                   & 130                    & 5700                     \\
Russian           & 1252                   & 9543918                  & 564                    & 1476045                  & 564                    & 25201                    \\
Spanish           & 648                    & 10373953                 & 369                    & 1669181                  & 369                    & 17749                    \\
Swahili           & 482                    & 131924                   & 239                    & 47793                    & 239                    & 7166                     \\
Telugu            & 828                    & 518079                   & 480                    & 66353                    & 480                    & 15429                    \\
Thai              & 733                    & 542166                   & 451                    & 128179                   & 451                    & 16313                    \\
Yoruba            & 119                    & 49043                    & 95                     & 33094                    & 95                     & 3022  \\ \hline

Average            & 750                    & 5907342                    & 439                     & 1088551                    & 439                     & 18819  \\           
\end{tabular}
\label{tab:miracl_stats}
\end{table*}

\begin{table*}[htb]
\caption{Comparison of the characteristics of MIRACL-VISION with vidore and vdr-multilingual benchmarks.}
\begin{tabular}{p{5cm}|p{2.5cm}|p{4cm}|p{4cm}}
                                     & \textbf{vidore}                                      & \textbf{vdr-multilingual}                  & \textbf{MIRACL-VISION}             \\ \hline
\# of different languages            & \multicolumn{1}{r|}{2}                                & \multicolumn{1}{r|}{5}                      & \multicolumn{1}{r}{18}             \\
\# of datasets                       & \multicolumn{1}{r|}{10}                               & \multicolumn{1}{r|}{5}                      & \multicolumn{1}{r}{18}             \\
avg. number of queries per dataset   & \multicolumn{1}{r|}{380}                              & \multicolumn{1}{r|}{300}                    & \multicolumn{1}{r}{483}            \\
avg. number of documents per dataset & \multicolumn{1}{r|}{672}                              & \multicolumn{1}{r|}{3000}                   & \multicolumn{1}{r}{18500}          \\
document selection                   & random                                               & random                                     & hard-negatives sampled from large corpus \\
modalities                           & text, charts, tables, infographics.                     & text, visual.                               & text                               \\
query generation                     & human-generated and synthetic with manual evaluation & synthetic generated with manual evaluation & human-generated                   
\end{tabular}
\label{tab:comparision_benchmarks}
\end{table*}


\section{Main results and discussion}

We present in this section experiments results comparing MIRACL and derived datasets, MIRACL-VISION and other visual document retrieval datasets.

\subsection{Retrieval accuracy on MIRACL and our derived text datasets}

\begin{table}[htb]
\caption{NDCG@10 of text embedding models on MIRACL original and our variants}
\begin{tabular}{lp{1.0cm}p{1.0cm}p{1.4cm}p{1.2cm}}
                              & MIRACL & MIRACL-1stPara-graph & MIRACL-1stParagraph-Reduced & MIRACL-VISION-text \\ \hline
Llama-3.2-1B (internal)                      & 0.6225 & 0.8231     & 0.8292         & 0.7932         \\   
gte-multilingual-base         & 0.6210 & 0.8072   & 0.8136           & 0.7682         \\
multilingual-e5-large         & 0.6512 & 0.8322  & 0.8323            & 0.7624         \\
arctic-embed-l-v2.0 & 0.6493 & 0.8289     & 0.8310         & 0.7806         \\

bge-m3                        & 0.6776 & 0.8442  & 0.8468            & 0.7964         \\ \hline    

Average                       & 0.6499 & 0.8271     & 0.8306         & 0.7798         \\                
\end{tabular}
\label{tab:miracl_text_recall}
\end{table}

In this section, we compare multilingual text embedding models - \textit{dse-qwen2-2b-mrl-v1}\footnote{https://huggingface.co/MrLight/dse-qwen2-2b-mrl-v1}\cite{ma2024unifying}, \textit{gme-Qwen2-VL-2B-Instruct}\footnote{https://huggingface.co/Alibaba-NLP/gme-Qwen2-VL-2B-Instruct}\cite{zhang2024gme},  \textit{vdr-2b-multi-v1}\footnote{https://huggingface.co/llamaindex/vdr-2b-multi-v1}, and \textit{colqwen2-v1.0}\footnote{https://huggingface.co/vidore/colqwen2-v1.0}\cite{faysse2024colpaliefficientdocumentretrieval} - on original MIRACL and our intermediate MIRACL text variants that are compatible with MIRACL-VISION: \textbf{MIRACL-1stParagraph-Reduced} and \textbf{MIRACL-VISION-text}, described in Section~\ref{sec:extension}. We calculated all scores for every model and dataset ourselves.

The selected multilingual embedding models were trained on the original MIRACL train split. 
For a fair comparison with the vision models, we fine-tune Llama 3.2 1B as an embedding model with constrastive loss and some data excluding MIRACL train set as a baseline model. Its retrieval accuracy is not much smaller than the other multilingual text embedding models though. 

As can be seen in Table~\ref{tab:miracl_text_recall}, the average NDCG@10 over all models is 0.6499 for the original MIRACL dataset and  0.8271 for our MIRACL-1stParagraph. 
It indicates that our MIRACL-1stParagraph-Reduced filtered version is easier than MIRACL. One hypothesis is that questions related to the first paragraph are easier or that other chunks from the same Wikipedia article (which we remove) are more challenging negatives. 

By comparing MIRACL-1stParagraph and MIRACL-1stParagraph-Reduced columns, we can notice they are close. That indicates that our method for reducing dataset size (by ~58x) while keeping hard negatives for questions is successful in maintaining retrieval accuracy correlation.

We also evaluate the models on \textbf{MIRACL-MIRACL-text}, which is the MIRACL-VISION version with text extracted from HTML that roughly matches the textual content present in the 1st page of the Wikipedia article. The average score decreases by 3 percent points to 0.7798. As extracted HTML text is longer than the original MIRACL chunked text from 1st paragraph, we believe the additional noise might make it more challenging for the models to retrieve the right content. Overall, the \textbf{MIRACL-VISION-text} behaves similarly to the filtered \textbf{MIRACL-1stParagraph} data.

\subsection{Retrieval accuracy of visual document retrieval datasets}
\label{sec:vlms_benchmark}

We compare MIRACL-VISION to other visual document retrieval benchmarks - ViDoRe and vdr-multilingual - using 4 public VLM-based embedding models, as shown in Table~\ref{tab:visual_benchmark_results}. The model with best average NDCG@10 - \textit{colqwen2-v1.0} - scores  0.9604 for \textit{vdr-multilingual} and 0.8969 for \textit{vidore benchmark}. Both benchmarks are almost saturated. One hypothesis could be that their small corpus size, with less than 3000 documents, is not challenging for retrieval. Another possibility is that generating synthetic questions with VLMs or LLMs have the tendencies to repeat phrases and keywords in the queries, making it easier to retrieve the right chunks. These synthetic questions may differ from real user-generated open questions, which are not biased toward rephrasing fragments or key words of the documents they seek to retrieve. 
The visual retriever models have significantly lower NDCG@10 in MIRACL-VISION (average 0.4715) than on other datasets, indicating it provides a challenging benchmark for multilingual visual document retrieval.

\begin{table}[htb]
\caption{NDCG@10 of VLM-based embedding models on visual document retrieval benchmarks}
\begin{tabular}{lp{1.4cm}ll}
                         & vdr-multilingual & vidore & MIRACL-Vision \\ \hline
dse-qwen2-2b-mrl-v1      & 0.8363           & 0.8416 & 0.4426        \\
gme-Qwen2-VL-2B-Inst & 0.9165           & 0.8878 & 0.5283        \\
vdr-2b-multi-v1          & 0.9371           & 0.8584 & 0.4741        \\
colqwen2-v1.0            & 0.9604           & 0.8969 & 0.4728        \\ \hline
Average                  & 0.9126           & 0.8712 & 0.4795        \\

\end{tabular}
\label{tab:visual_benchmark_results}
\end{table}

\subsection{How visual embedding models compare with text embedding models on multilingual document retrieval?}

In Table~\ref{tab:miracl_vision}, we compare the VLM-based embedding models on MIRACL-VISION with text embedding models on \textbf{MIRACL-VISION-text}, as both datasets contain the same questions and documents (represented as a document screenshot or text). The best vision model is \textit{gme-Qwen2-VL-2B-Instruct} with an average NDCG@10 score of 0.5283 and best public text-model is \textit{bge-m3} with 0.7964, outperforming the visual pipelines by over 50\%, as seen in Table~\ref{tab:miracl_vision}. A detailed analysis shows that the vision models do not work for Thelugu, with NDCG@10 below < 0.1. After removing the language as an outlier, the text-based models perform 43\% higher in average compared to visual embedding models. 

Table~\ref{tab:miracl_vision} provides the performance per language and we can see that the text-based versions perform better for every language. In case of English MIRACL, the gap is the smallest but still significant with 12.1\%. Common languages, such as Chinese, French or Spanish, have a similar pattern with up to 16.5\%. Arabic, Hindi and Thai, which are less common in research, wih non-Latin alphabet, have a performance gap of up to 59.7\%.

\begin{table*}[ht]
\caption{NDCG@10 of text embedding models and visual embedding models on MIRACL-VISION.}
\begin{tabular}{p{2.8cm}|p{1.2cm}p{1.2cm}p{1.2cm}p{1.2cm}p{1.2cm}|p{1.2cm}p{1.2cm}p{1.2cm}p{1.2cm}}
                               & \multicolumn{5}{c|}{\textbf{MIRACL-VISION (Text)}}                                                                                                                                                                                                  & \multicolumn{4}{c}{\textbf{MIRACL-VISION (Image)}}                                                                                                                                                   \\ \hline
                               & \textbf{multil-ingual-e5-large} & \textbf{arctic-embed-l-v2.0} & \textbf{gte-multilingual-base} & \textbf{bge-m3} & \textbf{Llama-3.2-1B (internal)} & \textbf{dse-qwen2-2b-mrl-v1} & \textbf{gme-Qwen2-VL-2B-Instruct} & \textbf{vdr-2b-multi-v1} & \textbf{colqwen2-v1.0} \\
\textbf{\# Params (in M)} & 560                                                & 567                                                        & 305                                                & 567                                 & 1235                                   & 1543                                             & 1543                                                  & 1543                                         & 1543                                       \\
\textbf{Language}              & \multicolumn{1}{l}{}                               & \multicolumn{1}{l}{}                                       & \multicolumn{1}{l}{}                               & \multicolumn{1}{l}{}                & \multicolumn{1}{l|}{}                  & \multicolumn{1}{l}{}                             & \multicolumn{1}{l}{}                                  & \multicolumn{1}{l}{}                         &                        \\ \hline
Arabic                         & 0.8557                                             & 0.8754                                                     & 0.8503                                             & 0.8883                              & 0.8833                                 & 0.3893                                           & 0.4888                                                & 0.4379                                       & 0.4129                                     \\
Bengali                        & 0.8421                                             & 0.8325                                                     & 0.8211                                             & 0.8585                              & 0.7902                                 & 0.2352                                           & 0.3755                                                & 0.2473                                       & 0.2888                                     \\
Chinese                        & 0.6900                                             & 0.7179                                                     & 0.7167                                             & 0.7458                              & 0.7561                                 & 0.5962                                           & 0.6314                                                & 0.5963                                       & 0.4926                                     \\
English                        & 0.7029                                             & 0.7437                                                     & 0.7345                                             & 0.7348                              & 0.7721                                 & 0.6605                                           & 0.6784                                                & 0.6784                                       & 0.6417                                     \\
Farsi                          & 0.6793                                             & 0.7001                                                     & 0.6984                                             & 0.7297                              & 0.7192                                 & 0.2250                                           & 0.3085                                                & 0.2398                                       & 0.2616                                     \\
Finnish & 0.8974 & 0.9014 & 0.8957 & 0.9071 & 0.9097 & 0.4162 &0.6863 & 0.5283 & 0.6604 \\
French                         & 0.7208                                             & 0.8236                                                     & 0.7771                                             & 0.8158                              & 0.8545                                 & 0.7160                                           & 0.6851                                                & 0.7194                                       & 0.6876                                     \\
German	& 0.7622 & 0.7774 & 0.7498 & 0.7695 & 0.7823 & 0.6267 & 0.6345 &	0.6205 & 0.5995 \\
Hindi                          & 0.7595                                             & 0.7255                                                     & 0.6916                                             & 0.7581                              & 0.7770                                 & 0.1740                                           & 0.3127                                                & 0.2058                                       & 0.2209                                     \\
Indonesian                     & 0.6793                                             & 0.6906                                                     & 0.6757                                             & 0.7049                              & 0.6977                                 & 0.4866                                           & 0.5416                                                & 0.5254                                       & 0.5320                                     \\
Japanese                       & 0.8378                                             & 0.8484                                                     & 0.8442                                             & 0.8720                              & 0.8802                                 & 0.6232                                           & 0.7305                                                & 0.6553                                       & 0.6970                                     \\
Korean                         & 0.7327                                             & 0.7545                                                     & 0.7397                                             & 0.7934                              & 0.8088                                 & 0.4446                                           & 0.6202                                                & 0.4952                                       & 0.4419                                     \\
Russian &  0.7857 &  0.8242 &  0.8023 &  0.8363 &  0.8468 &  0.6505 &  0.7202 & 
 0.6995 &  0.6811 \\
Spanish                        & 0.6596                                             & 0.7250                                                     & 0.7029                                             & 0.7268                              & 0.7318                                 & 0.5927                                           & 0.6277                                                & 0.6274                                       & 0.6224                                     \\
Swahili                        & 0.8157                                             & 0.8089                                                     & 0.7987                                             & 0.8337                              & 0.8059                                 & 0.4156                                           & 0.5348                                                & 0.4509                                       & 0.4931                                     \\
Telugu                         & 0.8948                                             & 0.9201                                                     & 0.9076                                             & 0.9090                              & 0.8101                                 & 0.0274                                           & 0.0893                                                & 0.0318                                       & 0.0264                                     \\
Thai                           & 0.8424                                             & 0.8485                                                     & 0.8509                                             & 0.8682                              & 0.8673                                 & 0.2692                                           & 0.3563                                                & 0.3177                                       & 0.2389               \\
Yoruba                         & 0.5655                                             & 0.5332                                                     & 0.5698                                             & 0.5842                              & 0.5839                                 & 0.4178                                           & 0.4884                                                & 0.4577                                       & 0.5120                                    
                             \\ \hline
Average	& 0.7624 & 0.7806 & 0.7682 & 0.7964 & 0.7932 & 0.4426 & 0.5283 & 0.4741 & 0.4728 \\
Average w/o Thelugu	& 0.7546 & 0.7724 & 0.7600 & 0.7898 & 0.7922 & 0.4670 & 0.5542 & 0.5002 &	0.4991 \\

\end{tabular}
\label{tab:miracl_vision}
\end{table*}

Table~\ref{tab:miracl_vision} also shows the number of parameters per model. In the case of VLMs, it includes only the LLM without the vision backbone. The \textit{gte-multilingual-base} outperforms the vision models with 5x less parameters (305M parameters vs. 1543M parameters of Qwen2-based VLMs model).

As an additional note, \textit{vdr-2b-multi-v1} is a continuous fine-tuning of \textit{dse-qwen2-2b-mrl-v1} based on \textit{vdr-multilingual-train} and reports significant gains on French and German on \textit{vdr-multilingual-test}. Overall, \textit{vdr-2b-multi-v1} has a better performance than \textit{dse-qwen2-2b-mrl-v1} on MIRACL-VISION, but we do not observe similar gains, which might indicate that the Qwen2-based models have multilingual capabilities but additional fine-tuning learns the data distribution of \textit{vdr-multilingual-test}.

\section{Limitations}

In this section, we discuss limitations and future research directions.

\subsubsection{Questions only about text modality} 
As MIRACL is a text-based dataset, most user queries are answered by text paragraph and MIRACL-VISION's modality is mainly text. The other visual document retrieval benchmark - ViDoRe and vdr-multilingual - provide questions backed by other modalities, such as charts, infographics or tables. However, we believe that the text modality is sufficient to evaluate multilingual capabilities of vision-based retrievers and our experiments demonstrate blindspots of current state-of-the-art models.

\subsubsection{MIRACL-VISION-text could be refined to match perfectly MIRACL-VISION textual content} 
Most visual document retrieval pipelines assume PDFs as input. Generating an image per PDF page is easy, but extracting text can be more challenging. The comparison of MIRACL-VISION-text with vision-based MIRACL-VISION relies on text extraction from the HTML body of each article. The extracted text is clean and might have a higher quality as extracted text from PDF. One option is to use an OCR pipeline to convert PDFs to image to text, but as PDFs can contain the text as input, it can be a mixed solution. Comparing different OCR pipelines is beyond the scope of this paper and the extracted HTML text is an upper bound for high-quality PDF to text conversion. 

\section{Conclusion}

In this paper, we introduced MIRACL-VISION, a multilingual benchmark for visual document retrieval. We described the methodology to generate MIRACL-VISION from MIRACL and Wikipedia articles, covering  18 different languages. We outlined our method for reducing the large corpus size by strategically selecting hard negatives. The resulting multilingual datasets are both challenging and efficiently sized, ensuring manageable computational requirements for evaluation. 

Our experiments demonstrate that current state-of-the-art vision embedding models on text-heavy pages have a lower retrieval accuracy compared to smaller text embedding models, across all languages. The performance gap is up to 59.7\%. Although some prior work suggests that the VLMs have zero-shot multilingual capabilities and that VLM-based document retrieval is superior to a text-based pipeline, MIRACL-VISION challenges the approach. We believe  the release of MIRACL-VISION will enable the community to gauge their progress towards more robust multilingual vision embedding models.  

In the future, we plan to provide a MIRACL-VISION train split and fine-tune visual embedding models on it. We also suggest enriching MIRACL-VISION with more modalities in multiple languages for multimodal multilingual evaluation. 

\bibliographystyle{ACM-Reference-Format}
\bibliography{manuscript}

\appendix

\section{Additional examples}
\label{sec:more_examples}

In this Appendix we provide additional examples of questions and corresponding document pages from MIRACL-VISION.

\begin{figure}[hb]
    \centering
    \includegraphics[width=1.0\linewidth]{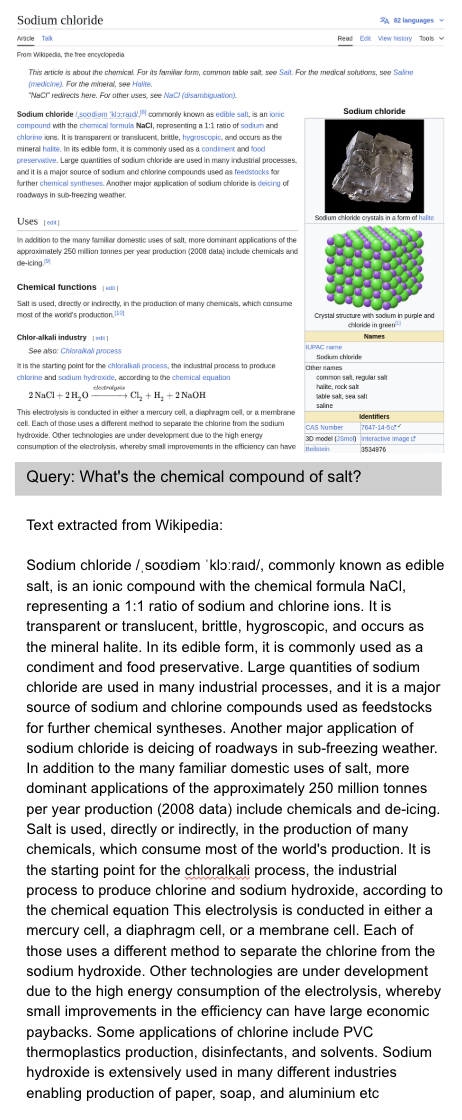}
    \caption{Example of a User Query and Document Image of MIRACL Vision. The below textbox is the extracted wikipedia text from the article.}
    \label{fig:example_1_withtext}
\end{figure}

\begin{figure}[hb]
    \centering
    \includegraphics[width=1.0\linewidth]{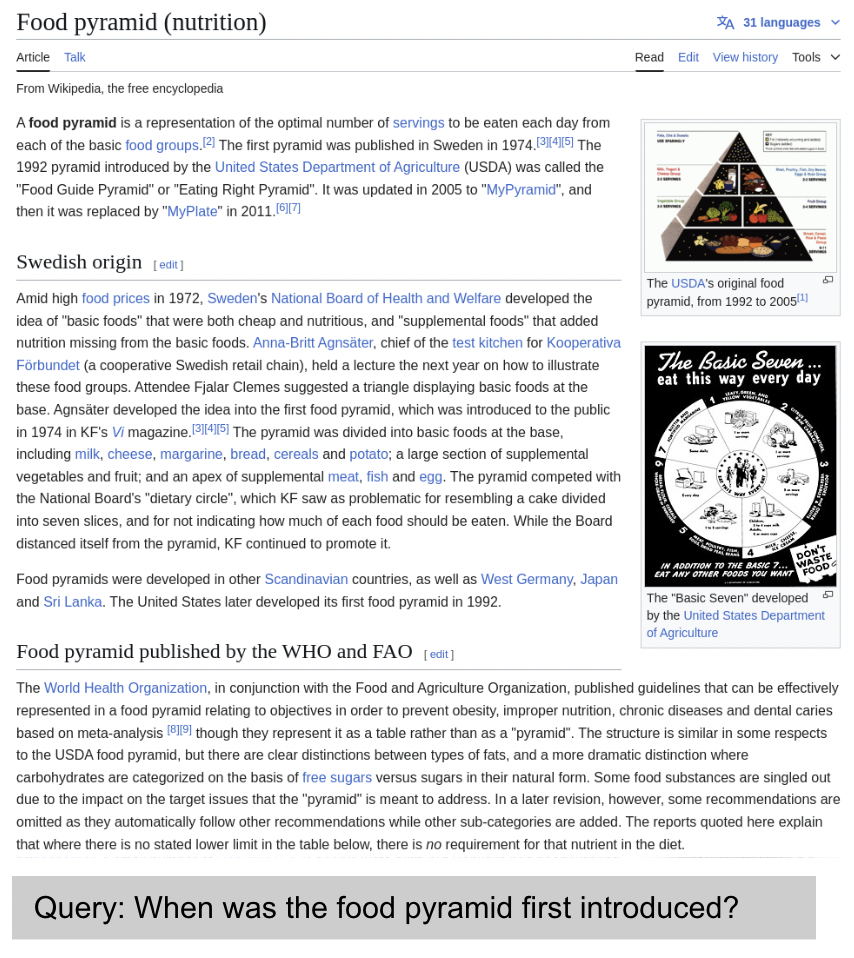}
    \caption{Example of a User Query and Document Image of MIRACL Vision}
    \label{fig:example_2}
\end{figure}


\end{document}